\documentclass[prd,amsmath,amssymb,superscriptaddress,twocolumn,nofootinbib,floatfix,showpacs]{revtex4}
\input{epsf}
\usepackage{epsf}
\usepackage{graphicx,epsfig}
\usepackage{bm}
\usepackage{latexsym,float}
\usepackage{graphicx}
\usepackage{bm}
\usepackage{latexsym,amssymb,amsmath,float,url}

\newcommand {\ga} {\ {\raise-.5ex\hbox{$\buildrel>\over\sim$}}\ }
\newcommand {\la} {\ {\raise-.5ex\hbox{$\buildrel<\over\sim$}}\ }

\def\bs{\begin{split}}
\def\es{\end{split}}

\def\5{\overline 5}

\newcommand{\beq}[1]{\begin{equation}\label{#1}}
\newcommand{\eeq}{\end{equation}}
\newcommand{\bea}[1]{\begin{eqnarray}\label{#1}}
\newcommand{\eea}{\end{eqnarray}}

\def\be{\begin{equation}}
\def\ee{\end{equation}}
\def\ba{\begin{eqnarray}}
\def\ea{\end{eqnarray}}

\def\be{\begin{equation}}
\def\ee{\end{equation}}
\def\ba{\begin{eqnarray}}
\def\ea{\end{eqnarray}}

\renewcommand{\Psi}{\varPsi}

\begin{document}

\title{Thermal Relic Abundances of Particles with Velocity-Dependent Interactions}  

\author{James~B.~Dent} \email{jbdent@asu.edu}
\affiliation{Department of Physics and School of Earth and Space
Exploration, Arizona State University, Tempe, AZ 85287-1404}

\author{Sourish~Dutta}
\email{sourish.d@gmail.com}
\affiliation{Department of Physics and Astronomy, Vanderbilt University,
Nashville, TN 37235}

\author{Robert~J.~Scherrer}
\email{robert.scherrer@vanderbilt.edu}
\affiliation{Department of Physics and Astronomy, Vanderbilt University,
Nashville, TN 37235}

\begin{abstract}
We reexamine the evolution of thermal relic particle abundances for the case
where the interaction rate depends on the particle velocities.
For the case of Sommerfeld enhancement, we show that the standard analytic
approximation, modified in a straightforward way, provides an estimate
of the relic particle abundance that is accurate
to within 10\% (in comparison to $<$ 1 \% error for the non-Sommerfeld-enhanced
case).  We examine the effect of kinetic decoupling
on relic particle abundances when the interaction rate depends on the velocity.
For the case of pure $p$-wave annihilation, the effect of kinetic decoupling
is an increase in the relic abundance, but the effect is negligible when the
kinetic decoupling temperature is much less than the chemical decoupling
temperature.  For the case of Sommerfeld-enhanced $s$-wave annihilations,
after kinetic decoupling occurs, annihilations continue to change the particle abundance
down to arbitrarily low temperatures, until either matter domination begins
or the Sommerfeld effect cuts off.  We derive analytic approximations to give
the final relic particle abundances for both of these cases.
\end{abstract}


\maketitle

\section{Introduction}

The calculation of the thermal evolution of particle abundances in the early universe
represents
one of the earliest and most fundamental applications of particle physics to cosmology 
\cite{Zel,Chiu,LW,Hut,Wolfram,Steigman,Olive,Krauss,ST,KT,Gondolo,Griest}.
The most important application of this calculation is the determination
of the relic dark matter abundance.  For the simplest case, that of $s$-wave
annihilation, one finds that the final abundance is essentially independent
of the mass, and the observed dark matter abundance can be achieved with
an annihilation rate of \cite{jungman}
approximately $\langle \sigma v \rangle \sim 3 \times 10^{-26}$ cm$^3$
sec$^{-1}$, where $\langle \sigma v \rangle$ is the thermally averaged product
of the cross section and relative velocity of the annihilating dark matter
particles.  This result is quite interesting, as it suggests that physics
at the electroweak scale may be responsible for the dark matter.

Recently an interesting twist has emerged in this calculation.  Motivated by a
desire to explain various anomalous astrophysical backgrounds, a
number of investigators have examined the possibility that dark matter
annihilation involves a Sommerfeld enhancement, which provides an additional
factor of $1/v$ in the dark matter annihilation cross section
\cite{Kamion,Arkani}.  The effect of this Sommerfeld enhancement on the thermal
relic abundances has been
discussed for specific models
in \cite{Baer,Hisano,Cirelli,March}, and treated more
generally by Kamionkowski and Profumo \cite{Kamion} and
Arkani-Hamed et al. \cite{Arkani}.  Note that the latter two papers
reached opposite conclusions regarding the effect of Sommerfeld-enhanced
annihilations on the relic abundance.  Kamionkowsk and Profumo estimated a 
significant suppression, while Arkani-Hamed et al. argued for a very small
effect.  However, these two conclusions are not actually inconsistent, because
they rely on different assumptions regarding the strength of the coupling
that induces the Sommerfeld enhancement.  We will examine both limiting
cases in our discussion below.

In this paper, we consider several new aspects of velocity dependent interactions, including
both Sommerfeld-enhanced $s$-wave annihilations, and pure $p$-wave annihilations.
In the next section, we first consider the analytic approximation of \cite{ST,KT},
modified for the case of Sommerfeld-enhanced $s$-wave annihilations, and show that it
provides a good approximation
to the relic abundances in this case.  We then consider the effects of kinetic decoupling,
which increases the rate at which the relic particle temperature declines, and thereby modifies
the abundance when the annilation rate has a velocity dependence.  We provide estimates
of this effect for both pure $p$-wave annihilation and Sommerfeld-enhanced $s$-wave annihilation.
The latter is a significantly larger effect; we find that
in this case, annihilations continue to decrease the relic abundance
down to arbitrarily late times, until
the abundance freezes out either at the onset of matter domination,
or when the Sommerfeld effect itself cuts off.  Although calculations of this sort can always
be done numerically for any particular model of a relic particle, it is useful to derive such analytic
estimates, since they can be applied to arbitrary models, and can provide qualitative insight
into the behavior of such models.  While interest
in the Sommerfeld enhancement has been spurred by
recent astrophysical observations, our discussion here is intended to
be as general as possible.
Our results are discussed in Sec. III.

\section{Calculation of Relic Abundances}
\subsection{Sommerfeld-enhanced $s$-wave annihilation}

Recall first the standard formalism for thermal particle abundances in
the early universe \cite{ST,KT,Gondolo}.  Let $n$ be the number density of a relic
particle $\chi$, and $n_{eq}$ be its thermal equilibrium number density.  Then
\be
\label{ndot}
\frac{dn}{dt} + 3Hn = -\langle \sigma v \rangle (n^2 - n_{eq}^2),
\ee
where $H$ is the Hubble parameter.  To eliminate the expansion term, we express the number density in
terms of $Y \equiv n/s$, where $s$ is the total entropy of the universe, and
we change the indendent variable to $x = m/T$.  Further, following 
\cite{ST,KT}, we parametrize the cross-section as
\be
\label{sigma}
\langle \sigma v \rangle = \sigma_0 x^{-n},
\ee
where $n=0$ corresponds to $s$-wave annilation,
$n = 1$ for $p$-wave annihilation, and so on.  Note that  \cite{Gondolo}
provides a more sophisticated treatment of $\langle \sigma v\rangle$, but at the level
of accuracy we are interested in here, equation (\ref{sigma}) will be
sufficient.  For all of the specific cases examined here, we take
$\sigma_0 = 3 \times 10^{-26}$ cm$^3$ s$^{-1}$.
Then equation (\ref{ndot})
becomes \cite{ST,KT}
\be
\label{dydx}
\frac{dY}{dx} = -\lambda x^{-n-2}(Y^2 - Y_{eq}^2),
\ee
where the constant $\lambda$ is given by
\be
\lambda = \sqrt{\pi/45}({g_{*S}}/{g_{*}^{1/2}})m_{Pl}m_\chi\sigma_0,
\ee
with $m_{Pl}  = 1/\sqrt{G}$.  Here $g_{*}$ is the effective number of
relativistic degrees of freedom in the universe, defined by the requirement
that the energy density in relativistic particles is $\rho_R =
(\pi^2/30)g_*T^4$, while $g_{*s}$ is defined in terms of the entropy density
$s$ as $s = (2\pi^2/45)g_{*s}T^3$.  For the cases we examine here, it
is accurate to take $g_{*s} \approx g_{*}$, and $g_{*}$ is given by
\begin{eqnarray}
g_* &=& 106.75~~~T > 175 {\rm ~GeV}\\
g_* &=& 96.25~~~175 {\rm ~GeV} > T > 80 {\rm ~GeV}\\
g_* &=& 86.25~~~80 {\rm ~GeV} > T > 4 {\rm ~GeV}\\
g_* &=& 75.75~~~4 {\rm ~GeV} > T > 150 {\rm ~MeV}\\
g_* &=& 17.25~~~150 {\rm ~MeV} > T > 20 {\rm ~MeV}\\
g_* &=& 10.75~~~T < 20 {\rm ~MeV}
\end{eqnarray}

For the case of interest here, the relic
particles can assumed to be nonrelativistic, so that $Y_{eq}$ is well-approximated
by Maxwell-Boltzmann statistics:
\be
Y_{EQ} = .145(g_\chi/g_{*})x^{3/2}e^{-x} \equiv ax^{3/2}e^{-x},
\ee
where $g_\chi$ is the number of degrees of freedom of the $\chi$ particle.

At early times, the relic particle is in thermal equilibrium, so that its
abundance tracks the equilibrium abundance, but at late times the abundance
freezes out to a constant value.  This argument can be made more explicit by
defining the quantity $\Delta \equiv Y-Y_{eq}$, the
evolution of which is given by
\be
\label{Delta}
\frac{d\Delta}{dx} = - \frac{dY_{eq}}{dx} - \lambda
x^{-n-2}\Delta(2Y_{eq}+\Delta).
\ee
The approximation in \cite{ST}
and \cite{KT} amounts to setting the right-hand side of equation (\ref{Delta})
to zero up to $x_f$, the value of $x$ at which the abundance freezes out, and
then integrating equation (\ref{Delta}) for $x > x_f$ with the assumption that
both $Y_{eq}$ and $dY_{eq}/dx$ are negligible.  One then obtains \cite{ST,KT}
\be
\label{xf}
x_f = ln[(n+1)a\lambda] - (n+1/2)ln[ln[(n+1)a\lambda]],
\ee
and the final value of $Y$ is
\be
\label{Yinf}
Y_\infty = \frac{3.79 (n+1) x_f^{n+1}}{(g_{*s}/g_*^{1/2})m_{Pl}m_\chi \sigma_0}.
\ee
This approximation agrees with the exact integration of the Boltzmann equation
to within a few percent.  For the case of $s$-wave annihilations, the evolution of
$\Delta$ is compared to the approximate evolution in Fig. 1 for a 500 GeV
particle.

Now consider what happens for $s$-wave annihilations that are Sommerfeld enhanced.
Sommerfeld enhancement arises from a long-range attractive force due to a light
force carrier $\phi$.  In the limit where $m_\phi \rightarrow 0$, the
annihilation cross-section is enhanced by the factor \cite{Arkani}
\be
\label{Seq}
S =  \frac{\pi \alpha /v}{1 - e^{-\pi \alpha/v}},
\ee
where $v$ is the velocity of the annihilating particles, and $4 \pi \alpha$ is
the square of the coupling.

Clearly, the magnitude of the Sommerfeld enhancement depends on the value
of $\alpha$.  We illustrate this effect in Fig. 1, showing how the evolution
of the particle abundance depends on $\alpha$.
\begin{figure}[t]
	\epsfig{file=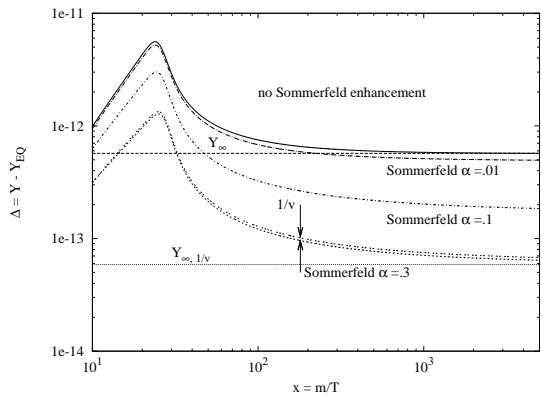,height=3.0in,angle=270}
	\caption
	{The evolution of relic particle densities for the case of
	Sommerfeld-enhanced $s$-wave
	annihilations of a 500 GeV mass particle 
	with $\sigma_0 = 3 \times 10^{-26}$ cm$^3$ s$^{-1}$ as a function of the indicated value of the
	coupling $\alpha$.  Top and bottom
	curves correspond to the limiting cases of no Sommerfeld enhancement,
	and $1/v$ enhancement.  Horizontal lines are the analytic estimates for
	the final relic abundances in these two cases (i.e. $n=0$ and $n=-1/2$,
	respectively, in equations \ref{xf} and \ref{Yinf}).
	}
\end{figure}
Clearly, for $\alpha \la 0.01$, the effect of the Sommerfeld enhancement on
the final relic particle abundance is negligible.  On the other hand,
for $\pi \alpha/v \gg 1$, equation (\ref{Seq}) reduces to a $1/v$
enhancement in the annihilation cross-section.  Fig. 1 shows that this limit
is achieved for $\alpha \ga 0.3$.  Note that \cite{Kamion} and
\cite{Arkani} assumed these opposite limiting behaviors in deriving their
estimates of the effect on the thermal relic abundance.
The case $0.01 \ga \alpha \ga 0.3$, gives
an intermediate regime displayed in Fig. 1.
Since $1/v$ enhancement provides one set of limiting behaviors, we will assume a simple
$1/v$ enhancement in what follows.  

Note that the effect is
more complex if one does not assume $m_\phi \rightarrow 0$; in this case, the
production of bound states results in resonant enhancement of the
annihilation rate,
while at the same time the Sommerfeld enhancement cuts off for $v <
m_\phi/m_\chi$ \cite{Arkani}.
We will consider only the case $m_\phi \rightarrow 0$, but will
discuss these other effects qualitatively later.

A nonrelativistic particle has a velocity that scales as
$\langle v^2 \rangle \propto T_\chi$.  As long as the particle is in thermal
equilibrium, $T_\chi = T$.  Hence the effect of Sommerfeld annihilation is
to modify equation (\ref{dydx}) (for $n=0$) to the form
\be
\frac{dY}{dx} = -\lambda x^{-3/2}(Y^2 - Y_{eq}^2).
\ee
It would appear, then, that the relic abundance in this case is
well-approximated by the standard freeze-out abundance for the case $n=-1/2$ in
Eqs. (\ref{xf}) and (\ref{Yinf}).  Indeed, this was the assumption made
in Ref. \cite{Kamion}.  However, it is not {\it a priori} obvious that
equations (\ref{xf}) and (\ref{Yinf}) can be accurately applied
in the regime $n < 0$, since they have been numerically tested only
in the regime $n > 0$, and the freeze-out process becomes progressively
less ``sharp" as $n$ decreases.  In
Fig. 1, we integrate the Boltzmann equation for the same set of parameters, but with
a $1/v$ enhancement in the annihilation rate.  The
final abundance is reasonably well-approximated by the $n=-1/2$ analytic
approximation, but the agreement with the exact numerical results is
not quite as good as for the s-wave case without the Sommerfeld effect.

This result allows us to estimate the ratio between the abundance in the
presence of Sommerfeld enhancement, $Y^{SOM}_\infty$ to the standard $s$-wave abundance
without Sommerfeld enhancement, $Y_\infty$.  We obtain
\begin{eqnarray}
\frac{Y^{SOM}_\infty}{Y_\infty} &=& \frac{1}{2}
\frac{x_{fSOM}^{1/2}}{x_{f}},\nonumber \\
\label{ratio}
&=& \frac{1}{2} \frac{\sqrt{\ln (a \lambda/2)}}
{\ln(a\lambda) - (1/2)\ln\ln(a\lambda)}.
\end{eqnarray}
Here $x_{fSOM}$ is the value of $x_f$ when Sommerfeld enhancement is included.
Taking $x_{fSOM} \approx x_{f0}$ yields the abundance estimate given
in \cite{Kamion}.  Our results confirm that the change in the
$x_f$ is indeed very small.  For the standard dark matter freeze-out value
of $x_f = 20$, we find that $x_{fSOM}$ is larger by only a few percent, while
equation (\ref{ratio}) gives $Y^{SOM}_\infty/Y_\infty \sim 1/10$.  Both of these results are confirmed
numerically in Fig. 1.
\begin{figure}[t]
	\epsfig{file=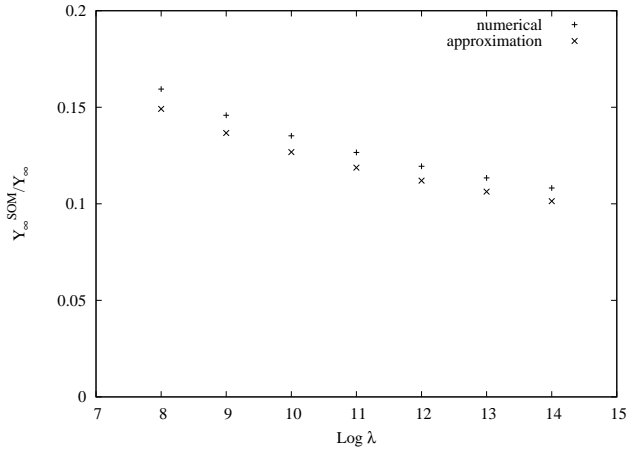,width=2.4in,angle=270}
	\caption
	{A comparison between the analytic approximation for the ratio in relic
	abundance for $s$-wave annihilation with and without a $1/v$ Sommerfeld
	enhancement to the corresponding numerical results over several orders of magnitude of 
	the parameter $\lambda$.
	}
\end{figure}
In Fig. 2, we compare the estimate given by equation (\ref{ratio}) to numerical
results.  The analytic estimate differs from the numerical result by
about 10\%.  In contrast, the analytic abundance estimate for s-wave
annihilation without Sommerfeld enhancement is accurate to within 1\% \cite{ST}.

\subsection{Effects of Kinetic Decoupling}

The results in the previous section assume that the temperature of the relic annihilating particle tracks
the background radiation temperature.  This will be true
as long as the annihilating particle remains in thermal equilibrium with
the radiation background.  However, once the particle
drops out of thermal equilibrium, we expect its temperature to
scale as $T_\chi \propto 1/R^2$, where $R$ is the
scale factor, while the radiation temperature scales as $T \propto 1/R$.

Thus, we need to make the standard distinction between chemical equilibrium and
kinetic equilibrium.  The freeze-out
process we have discussed above is actually the process by which the annihilating relic particle
drops out of chemical equilibrium, so that number-changing interactions
are no longer effective, and the particle's comoving number density becomes constant.
However, even after dropping out of chemical equilibrium, the relic particle will, in general remain in kinetic equilibrium as it 
continues to scatter off of relativistic standard model particles which are in local thermal equilibrium with the radiation background.
As long as the particle is in kinetic equilibrium, its temperature tracks the background radiation
temperature.  Finally, at some kinetic decoupling temperature, $T_k$, the scattering interactions
are no longer sufficient to maintain kinetic equilibrium, and the temperature of the particle
decreases as $1/R^2$ rather than $1/R$.  (For a recent discussion, see, e.g.,
\cite{Chen,Hofmann,BH,Bring}).

The precise temperature at which kinetic decoupling occurs is dependent on the
model
for the relic particle of interest.
For instance, in the set of supersymmetric
models examined in  \cite{Bring}, $T_k/T_f \sim 10^{-1} - 10^{-3}$.  Since we wish to keep
our discussion as general as possible, we will take $T_k/T_f$ as a free parameter, subject
only to the constraint that
$T_k \le T_f$, since number-changing interactions also
maintain kinetic equilibrium.  We also make the approximation that the particle
drops out of kinetic equilibrium
instantaneously at $T_k$; this is a reasonable approximation \cite{BH,Bring}.
With these assumptions, the relation between $T_{\chi}$ and $T$ is
\be
T_\chi = T^2/T_k.
\ee

The change in the evolution of $T_\chi$ brought about by kinetic decoupling changes the
velocity evolution of the annihilating particles, since $v \propto T_\chi^{1/2}$ for nonrelativistic
particles.  Since the standard $s$-wave annihilation cross section, $\langle \sigma v \rangle$,
is independent of $T_\chi$ (or equivalently, $v_\chi$), kinetic decoupling has no effect in
this case.  The same is not true for $p$-wave annihilation, for which $\langle \sigma v\rangle
\propto T_\chi$, or for Sommerfeld-enhanced $s$-wave annihilation, for which  $\langle \sigma v \rangle
\propto T_\chi^{-1/2}$.  The reverse reactions (which create $\chi$) can be neglected
during the era following kinetic decoupling, since $T_k \le T_f$.
Thus, the Boltzmann equation following kinetic
decoupling for
$p$-wave annihilation becomes
\be
\label{pkinetic}
\frac{dY}{dx} = -\lambda x_k x^{-4}Y^2,
\ee
where we define the constant $x_k = m/T_k$.
For Sommerfeld-enhanced $s$-wave annihilation, we obtain:
\be
\label{Skinetic}
\frac{dY}{dx} = -\lambda x_k^{-1/2} x^{-1}Y^2.
\ee

The effect of kinetic decoupling on the final relic abundances is easy to
estimate.  Recall that equation (\ref{Yinf}) is derived by
integrating the annihilation portion of the Boltzmann equation (only)
from $x = x_f$ to $\infty$ \cite{ST,KT}.  Replacing this integration
by an integration from $x_f$ to $x_k$, and then integrating
equations (\ref{pkinetic}) and (\ref{Skinetic})
from $x_k$ to $\infty$ should provide the correct estimate of the change
in the final relic abundance.

For $p$-wave annihilation, we obtain the ratio between the final abundance
in the presence of kinetic decoupling, $Y^{(k)}_\infty$, and the abundance
in the limit where the particle stays in kinetic equilibrium to an
arbitrarily low temperature, $Y_\infty$.  This ratio is
\be
\frac{Y^{(k)}_\infty}{Y_\infty} = \frac{1}{1- (1/3)(T_k/T_f)^2}.
\ee
We see that the effect of kinetic decoupling is to increase the final
relic abundance for the case of $p$-wave annihilations.  This easy to
understand, since the annihilation rate in this case scales as $T_\chi$,
so a more rapid decrease in $T_\chi$ due to kinetic decoupling leads to
fewer relic annihilations after freeze-out, and so a larger relic abundance.
The effect, however, rapidly becomes irrelevant for
$T_k/T_f << 1$.   For example, for $T_k/T_f = 1/2$, the result is a 9\%
increase in the relic abundance.  For $T_k/T_f < 0.1$, the increase
in the relic abundance is less than 0.3\%.
A numerical calculation of this effect is illustrated in Fig. 3.
\begin{figure}[t]
\epsfig{file=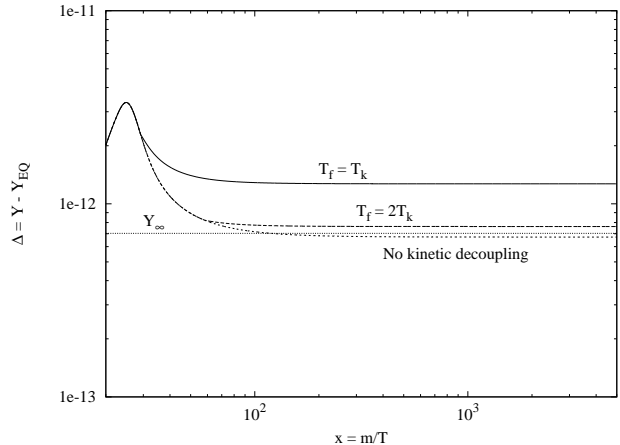,width=2.4in,angle=270}
\caption{The effect of kinetic decoupling on the evolution of the relic
particle abundance for the case of $p$-wave annihilation of a 500 GeV mass
particle with $\sigma_0 = 3 \times 10^{-26}$ cm$^3$ s$^{-1}$.
Horizontal line gives our analytic estimate of the final relic abundance.}
\end{figure}
Since the effect of kinetic decoupling becomes significant only
for values of $T_k/T_f$ that are implausibly large, it is unlikely to be of much
importance for p-wave annihilation.

The effect of kinetic decoupling is much more striking for the case of
Sommerfeld-enhanced $s$-wave annihilations.  In this case, an integration of
equation (\ref{Skinetic}) shows that annihilations never terminate: after
kinetic decoupling, $Y \sim 1/\ln x$.  (This case
has previously been discussed briefly in \cite{ST,Arkani}).
However, this process will eventually be cut
off by one of two possibilities.  First, as noted earlier, Sommerfeld enhancement
saturates once the velocity drops to $v \sim m_\phi/m_\chi$, at which point normal $s$-wave
annihilations resume.  Second, our calculation holds only for the
radiation-dominated case, and freeze-out will occur rapidly once matter
domination begins.  Let $T_{cutoff}$ be the radiation temperature at which
the Sommerfeld effect cuts off or
matter domination begins, whichever is larger.  Then we can again
integrate the equations governing particle annihilation from
$x_f$ to $x_k$ with $T_\chi = T$, and from $x_k$ to $x_{cutoff}$
with $T_\chi = T^2/T_k$, where freeze-out then occurs with negligible
further annihilations at $T_{cutoff}$.  We find
\be
\label{Yk}
\frac{Y^{(k)}_\infty}{Y_\infty} =  (T_f/T_k)^{1/2}(\sqrt{\frac{T_f}{T_k}} -1 + \frac{1}{2}\ln(T_k/T_{cutoff}))^{-1}.
\ee

The effect of
kinetic decoupling with Sommerfeld-enhanced annihilations is
illustrated numerically in Figs. 4 and 5. In Fig. 4, we show
the evolution of the particle abundance for the case we have just considered
($1/v$ enhancement), while Fig. 5 shows the case $\alpha = 0.01$ (of course,
our analytic estimate, equation (\ref{Yk}), does not apply
in the latter case.)  Fig. 5
illustrates the fact that a value of the coupling for
Sommerfeld enhancement can be small enough to produce
a negligible change in the relic abundance without kinetic decoupling, but it
can have a large effect once kinetic decoupling occurs.
\begin{figure}[t]
\epsfig{file=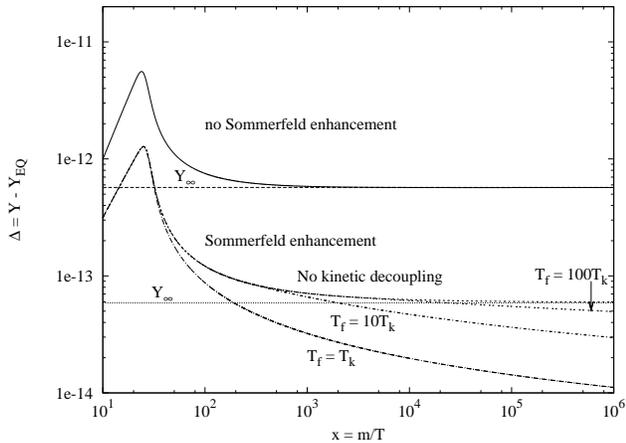,width=2.4in,angle=270}
\caption{The effect of kinetic decoupling on the evolution of the relic
particle abundance for the case of $s$-wave annihilation for a 500 GeV mass particle
with $\sigma_0 = 3 \times 10^{-26}$ cm$^3$ s$^{-1}$, in the limit
where the Sommerfeld enhancement scales as $1/v$.  Horizontal lines give our
analytic estimates of the final relic abundances.}
\end{figure}\begin{figure}[t]
\epsfig{file=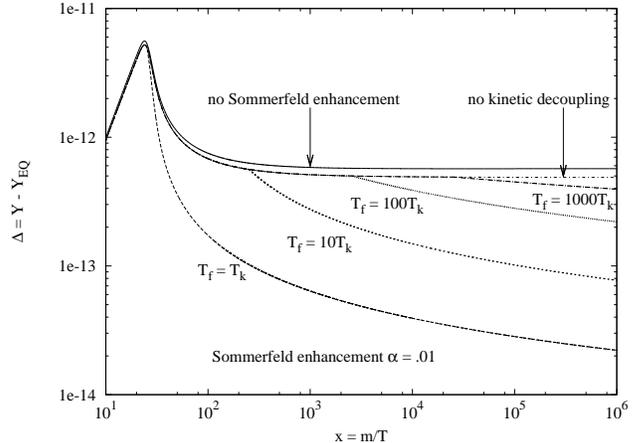,width=2.4in,angle=270}
\caption{As Fig. 4, for Sommerfeld-enhancement coupling of $\alpha = 0.01$, a value for which the Sommerfeld effect by itself is negligible without kinetic decoupling. Note the strong effect of kinetic decoupling upon the relic particle abundances. }
\end{figure}

\section{Discussion}

We have confirmed that the standard analytic approximation for the relic
particle abundances can be applied, with the appropriate modification, to the
case of $s$-wave relic abundances in the presence of a Sommerfeld enhanced
interaction, although the error in applying this approximation to
the case of Sommerfeld-enhanced s-wave annihilations ($\sim 10\%$)
is significantly larger than in the s-wave case without Sommerfeld
enhancement ($<$ 1\%).  We have also determined the range of the
coupling $\alpha$ over which Sommerfeld annihilation can be either neglected 
in the calculation of relic densities (as
suggested in  \cite{Arkani}) or treated purely as a $1/v$ enhancement
to the annihilation rate (as in \cite{Kamion}).

When kinetic decoupling occurs, it affects the relic abundances for both
$p$-wave annihilations and Sommerfeld-enhanced $s$-wave annihilations.
In the former case, the effect is generally very small unless kinetic decoupling occurs
at nearly the same epoch as chemical decoupling.  For Sommerfeld-enhanced
$s$-wave decoupling, the effect is quite large, and we have provided an
analytic estimate of this effect.

Finally, we note that another, quite different mechanism to produce a
velocity-dependent cross section is for a pole to lie near twice the mass of the
annihilating particle \cite{Griest}.  The effect is most striking when the pole
lies slightly below twice the particle mass \cite{Murayama}.  In this case,
just as for Sommerfeld-enhanced annihilation following kinetic decoupling,
the annihilations do not freeze out until the velocity drops below a cut-off
scale in the model.  Since the relic abundance in his model is set by this
cut-off scale, one would not expect a large change in the final
relic abundance if the annihilating particles
also kinetically decoupled.  However, a more detailed calculation likes outside
the scope of this paper.

\acknowledgments

J.B.D. , S.D. and R.J.S. were supported in part by the Department of
Energy (DE-FG05-85ER40226) at Vanderbilt.  J.B.D. also acknowledges support from a Department of Energy grant at Arizona State University and from the Arizona State Foundation.  We thank M. Kamionkowski, T.W. Kephart, and L.M. Krauss for helpful discussions.


\end{document}